# MedVKAN: Efficient Feature Extraction with Mamba and KAN for Medical Image Segmentation

Hancan Zhu[1], Jinhao Chen[1] and Guanghua He[1,*]

[1] School of Mathematics, Physics and Information, Shaoxing University, Shaoxing, Zhejiang, 312000, China

*Corresponding author:

Guanghua He, 900 ChengNan Rd, School of Mathematics Physics and Information, Shaoxing University, Shaoxing, Zhejiang, China 312000. Email: hegh23@sina.com

## Abstract

Medical image segmentation has traditionally relied on convolutional neural networks (CNNs) and Transformer-based models. CNNs, however, are constrained by limited receptive fields, while Transformers face scalability challenges due to quadratic computational complexity. To overcome these issues, recent studies have explored alternative architectures. The Mamba model, a selective state-space design, achieves near-linear complexity and effectively captures long-range dependencies. Its vision-oriented variant, the Visual State Space (VSS) model, extends these strengths to image feature learning. In parallel, the Kolmogorov-Arnold Network (KAN) enhances nonlinear expressiveness by replacing fixed activation functions with learnable ones. Motivated by these advances, we propose the VSS-Enhanced KAN (VKAN) module, which integrates VSS with the Expanded Field Convolutional KAN (EFC-KAN) as a replacement for Transformer modules, thereby strengthening feature extraction. We further embed VKAN into a U-Net framework, resulting in MedVKAN, an efficient medical image segmentation model. Extensive experiments on five public datasets demonstrate that MedVKAN achieves state-of-the-art performance on four datasets and ranks second on the remaining one. These results underscore the effectiveness of combining Mamba and KAN while introducing a novel and computationally efficient feature extraction framework. The source code is available at: https://github.com/beginner-cjh/MedVKAN.

**Keywords:** Medical Image Segmentation; Mamba; Kolmogorov-Arnold Network; Transformer

## 1 Introduction

Medical image segmentation plays a crucial role in modern clinical practice, with widespread applications in assisted diagnosis, treatment planning, and therapeutic outcome evaluation [1-3]. It facilitates precise identification and quantification of pathological regions, providing essential guidance for surgical planning and treatment monitoring. However, traditional segmentation methods rely heavily on manual annotation and expert interpretation, making them time-consuming, labor-intensive, and susceptible to inter-observer variability, which can compromise diagnostic consistency [4, 5]. To address these limitations, deep learning-based automated segmentation methods have emerged as a transformative technology, offering enhanced efficiency, accuracy, and reproducibility [6, 7].

Deep learning techniques, particularly convolutional neural networks (CNNs) and Transformer-based models, have led to significant advancements in medical image segmentation [8-11]. CNN-based architectures, such as U-Net [8], ResNet [12], nnU-Net [9], and SegResNet [13], leverage weight sharing and pooling mechanisms to efficiently extract local features, making them well-suited for processing medical images. However, due to the intrinsic limitations of convolutional kernels, CNNs struggle to capture global contextual information. Although dilated convolutions or deeper stacking can theoretically enlarge the receptive field, the effective receptive field observed in practice is often much smaller than the theoretical one. In contrast, Transformer-based models, including Swin Transformer [14], UNETR [15], SwinUNETR [16], employ multi-head self-attention (MHSA) mechanisms to capture global dependencies effectively. Nevertheless, the explicit computation of pairwise attention among all tokens results in quadratic computational complexity, substantially increasing the memory and computation costs for high-resolution medical images.

In recent years, state-space models (SSMs) [17, 18] have been widely applied in computer vision. Among them, Mamba [19], a recently proposed SSM, has attracted considerable attention due to its hardware-aware design and selective scanning mechanism, which enable strong global feature extraction with linear computational complexity, making it a promising alternative to Transformers. As a result, many studies have compared Mamba with Transformers, and it has demonstrated competitive performance in fields such as natural language processing and medical image analysis [20-24]. For instance, U-Mamba [21] introduced a hybrid module within the nnU-Net framework [9], integrating CNN's local feature extraction with Mamba's global modeling ability, marking an initial exploration of Mamba blocks in medical image analysis. The Visual State Space (VSS) model [22] enhanced Mamba by introducing the 2D Selective Scanning (SS2D) module, which scans images along four directions to capture richer contextual information. Swin-UMamba [25] extended prior Mamba-based architectures by incorporating ImageNet pretraining, thereby combining general-purpose pre-learned visual representations with Mamba's efficient global modeling capabilities. Additionally, SegMamba [24] proposed a three-directional spatial Mamba module, which unfolds images along three axes to improve 3D sequential modeling.

Meanwhile, Kolmogorov-Arnold Networks (KANs) [26, 27], grounded in the Kolmogorov-Arnold representation theorem [28, 29], have demonstrated exceptional nonlinear expressiveness by parameterizing B-spline functions. In tasks such as data fitting and solving partial differential equations, KANs significantly outperform multi-layer perceptrons (MLPs) while requiring substantially fewer parameters. The potential of KANs has also been extensively explored in computer vision. For instance, UKAN [30] incorporated Tok-KAN as a feature extractor within the U-Net framework, providing initial evidence of KAN's effectiveness in medical image segmentation. KAT [31] introduced GR-KAN, which replaces B-spline functions with rational functions and integrates KAN with Transformers, effectively mitigating KAN's computational inefficiencies. Conv-KAN [32] proposed a KAN-based convolutional layer, where convolutional kernels are parameterized using learnable B-spline interpolation functions, reducing parameter complexity while maintaining high accuracy. Furthermore, TransUKAN [33] introduced EfficientKAN, which integrates KAN with Transformers to leverage KAN's local nonlinear modeling capabilities, further optimizing Transformer architecture and performance.

Given the significant advantages of Mamba and KAN modules in feature extraction, prior studies have investigated replacing MHSA blocks with Mamba blocks [24] and substituting MLP blocks with KAN blocks [31, 33]. However, the integration of Mamba and KAN remains largely unexplored. Effectively leveraging their complementary strengths to enhance feature extraction and overall model performance remains a challenging research problem. To address this gap, we propose the VSS-Enhanced KAN (VKAN) module, which integrates VSS with the Expanded Field Convolutional KAN (EFC-KAN). We then integrate VKAN within a U-Net-inspired framework for medical image segmentation, termed MedVKAN. MedVKAN adopts a multi-stage encoder-decoder architecture with skip connections, designed to balance local and global feature representation. Specifically, convolutional layers are employed in the first three encoder stages to effectively capture local features, while the last two encoder stages incorporate the proposed VKAN blocks to extract global features, thereby achieving a favorable trade-off between segmentation accuracy and computational efficiency. The primary contributions of this study are as follows:

- We integrate Mamba and KAN into the VKAN block as a potential alternative to Transformer architecture, leveraging Mamba's global feature extraction and KAN's nonlinear expressiveness to enhance image feature extraction.
- To improve information interaction among adjacent pixels, we incorporate two additional 3×3 convolutions before the KAN module. This design effectively strengthens feature representation within the model.
- Extensive experiments on five public datasets demonstrate that MedVKAN achieves state-of-the-art performance on four datasets and ranks second on the remaining one, highlighting its effectiveness, robustness, and broad applicability in medical image segmentation.

## 2 Method

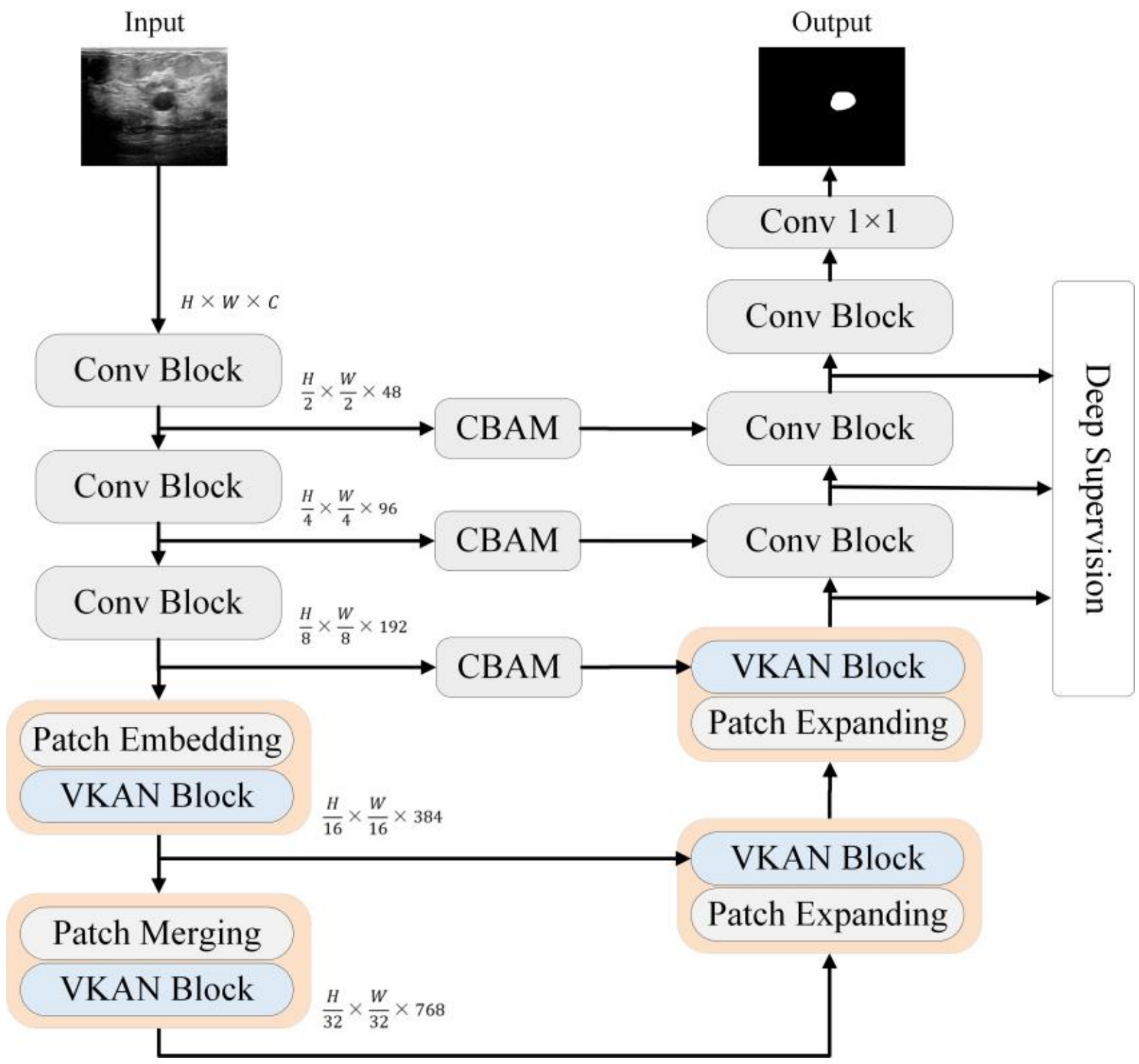

Figure 1. Schematic of the MedVKAN framework, comprising an encoder, decoder, and skip connections. It integrates convolutional blocks, VKAN blocks, and CBAM modules to enhance feature representation and segmentation performance.

As illustrated in Figure 1, the proposed MedVKAN model comprises three main components: the encoder, the decoder, and skip connections. The encoder consists of three convolutional blocks, two VKAN blocks specifically designed in this work, as well as Patch Embedding [11] and Patch Merging [14] operations. The decoder adopts a structure similar to the encoder, aiming to progressively restore the spatial resolution of feature maps. The skip connections integrate low-level features from the encoder with high-level abstract features from the decoder, preserving fine-grained details while restoring spatial resolution. To further enhance segmentation performance, a deep supervision mechanism is incorporated, enabling auxiliary predictions at multiple scales [34].

### 2.1 Feature Extraction in MedVKAN

Let $I \in \mathbb{R}^{B\times C\times H\times W}$ denote an input image, where $B$ is the batch size, $C$ the number of channels, and $H$ and $W$ the image height and width, respectively. The feature extraction process is divided into three main components: the encoder, skip connections, and the decoder, as described below.

In the encoder, the input image first passes through the initial convolutional block for preliminary feature extraction, producing a feature map $x_{e,1} \in \mathbb{R}^{B\times 48\times\frac{H}{2}\times\frac{W}{2}}$. Subsequently, the feature map is processed through the second and third convolutional blocks, extracting progressively deeper features and generating feature maps $x_{e,2} \in \mathbb{R}^{B\times 96\times\frac{H}{4}\times\frac{W}{4}}$ and $x_{e,3} \in \mathbb{R}^{B\times 192\times\frac{H}{8}\times\frac{W}{8}}$, respectively. Here, $x_{e,i}$ represents the output feature map at the $i$-th stage of the encoder. Each convolutional block comprises two consecutive 3×3 convolution operations followed by a max pooling (MP) operation for down-sampling. The computation process is formulated as follows:

$$\bar{x}_{e,i+1} = ReLU(BN\left(Conv(x_{e,i})\right),$$

$$x_{e,i+1} = MP\left(ReLU\left(BN\left(Conv(\bar{x}_{e,i+1})\right)\right)\right).$$

Here, Conv denotes a 3×3 convolution operation, BN stands for batch normalization, and ReLU refers to the Rectified Linear Unit activation function.

Next, the feature map $x_{e,3}$ undergoes downsampling via Patch Embedding, which applies a 2×2 convolution with a stride of 2 to reduce spatial resolution while increasing the channel dimension. The resulting feature map is then processed by a VKAN block, yielding $x_{e,4} \in \mathbb{R}^{B\times\frac{H}{16}\times\frac{W}{16}\times 384}$. Subsequently, Patch Merging partitions the feature map into four regions and concatenates them along the channel dimension. A linear projection then reduces the channel dimension to 768, followed by another VKAN block, generating the final encoder output $x_{e,5} \in \mathbb{R}^{B\times\frac{H}{32}\times\frac{W}{32}\times 768}$. The VKAN block is described in detail in Section 2.2.

In the skip connections, we introduce the CBAM module [35] to enhance the feature maps $x_{e,1}$, $x_{e,2}$, and $x_{e,3}$ from the first three encoder stages by refining salient information across both channel and spatial dimensions. The refined feature maps, denoted as $s_1, s_2$ and $s_3$, are then utilized for feature fusion in the decoding process. A detailed explanation of the CBAM module is provided in Section 2.3.

The decoder is designed symmetrically to the encoder. First, the encoded feature map $x_{e,5}$ is upsampled using Patch Expanding [36], where a linear projection initially maps it to $\mathbb{R}^{B\times\frac{H}{32}\times\frac{W}{32}\times 1536}$, followed by a rearrange operation that reshapes it into $\mathbb{R}^{B\times\frac{H}{16}\times\frac{W}{16}\times 384}$. The resulting feature map is then concatenated along the channel dimension with the feature map $x_{e,4}$ transmitted through the skip connection. To ensure compatibility, a 1×1 convolution restores the channel dimension to 384, followed by processing through a VKAN block, yielding $x_{d,5} \in \mathbb{R}^{B\times\frac{H}{16}\times\frac{W}{16}\times 384}$.

This process is iteratively applied, producing $x_{d,4} \in \mathbb{R}^{B\times\frac{H}{8}\times\frac{W}{8}\times 192}$. Next, three convolutional blocks further refine the features while progressively restoring spatial resolution, generating the feature maps $x_{d,3} \in \mathbb{R}^{B\times 96\times \frac{H}{4}\times \frac{W}{4}}, x_{d,2} \in \mathbb{R}^{B\times 48\times \frac{H}{2}\times \frac{W}{2}}$ and $x_{d,1} \in \mathbb{R}^{B\times 24\times H\times W}$. Finally, a 1×1 convolution performs channel compression, producing the final segmentation output.

### 2.2 VKAN Block

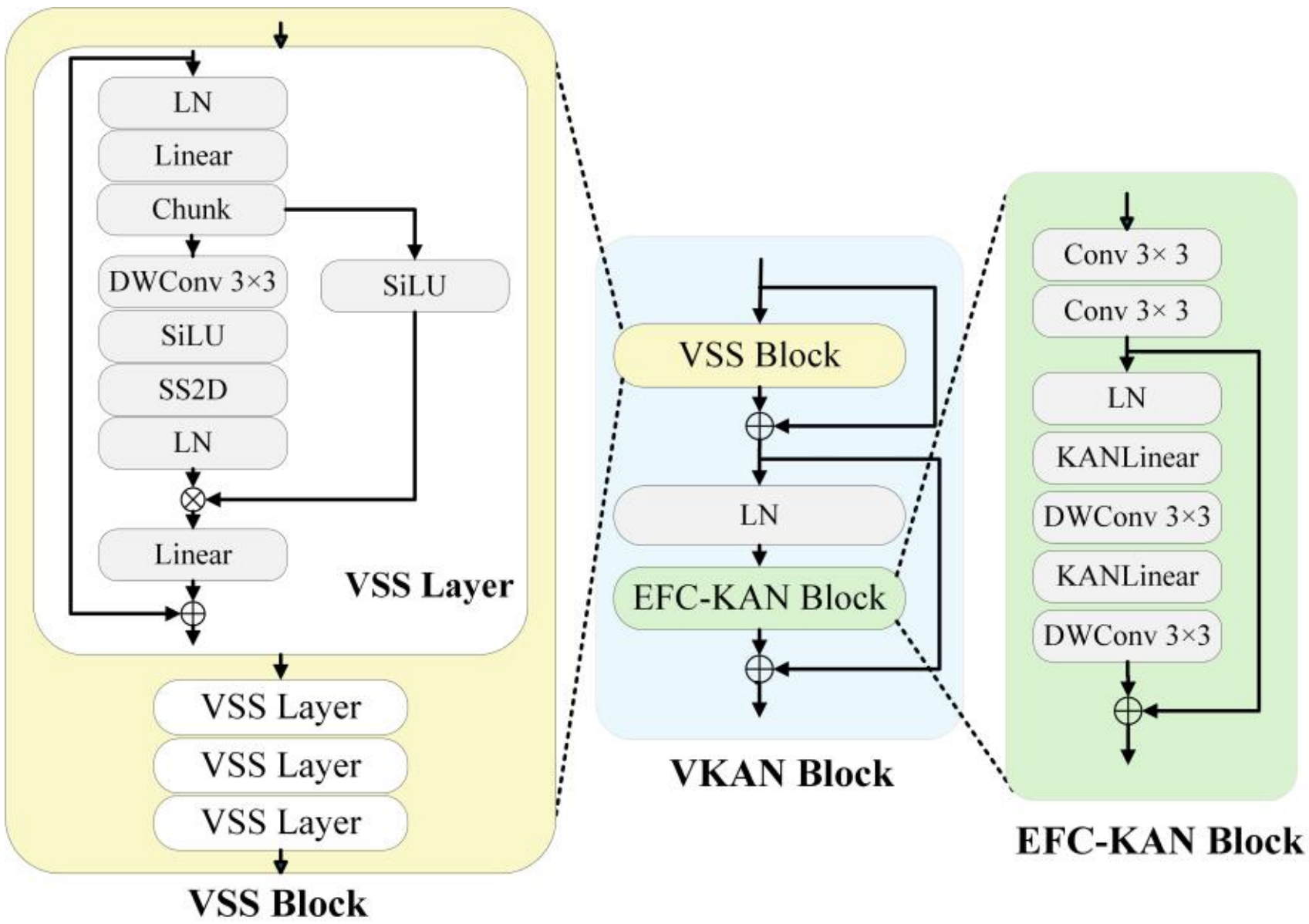

Figure 2. Structural diagram of the VKAN block. The input feature map is sequentially processed through the VSS block, Layer Normalization (LN), and the EFC-KAN block, with a residual connection facilitating element-wise addition to enhance feature preservation.

In MedVKAN, we introduce the VKAN block as a replacement for the Transformer block, integrating a VSS block and an EFC-KAN block to enhance global feature extraction and nonlinear representation while reducing computational complexity. As shown in Figure 2, the VKAN module first processes input features through the VSS block, followed by a residual connection and Layer Normalization (LN). The transformed features then pass through the EFC-KAN block, with an additional residual connection further refining the module's output.

#### 2.2.1 VSS Block

As shown in Figure 2, the VSS block consists of four VSS layers, an improved version of Mamba, each designed based on [22]. Given an input feature map $x \in \mathbb{R}^{B\times H\times W\times C}$, the process begins with LN, followed by a linear projection that expands the channel dimension to $4C$, producing $\bar{x} \in \mathbb{R}^{B\times H\times W\times 4C}$. The expanded feature map is then split along the channel dimension into two components, $\bar{x}_1, \bar{x}_2 \in \mathbb{R}^{B\times H\times W\times 2C}$, using a chunking operation.

For feature extraction, $\bar{x}_1$ is first permuted to $\mathbb{R}^{B\times 2C\times H\times W}$ and sequentially processed by depthwise convolution (DWConv), SiLU activation, and the SS2D module to enhance spatial feature representation. The output is then reshaped back to $\mathbb{R}^{B\times H\times W\times 2C}$ and normalized. Next, the refined $\bar{x}_1$ is element-wise multiplied with the SiLU-activated $\bar{x}_2$, yielding the updated feature representation $\bar{x} \in \mathbb{R}^{B\times H\times W\times 2C}$. To restore the original channel dimension, a linear projection maps $\bar{x}$ back to $\mathbb{R}^{B\times H\times W\times C}$. Finally, a residual connection adds $\bar{x}$ to the original input $x$, ensuring stable feature propagation and enhancing representational capacity.

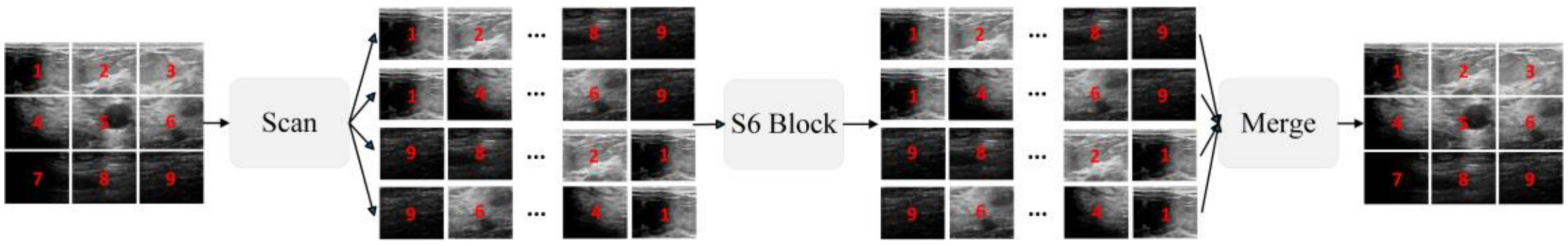

Figure 3. Schematic illustration of the 2D Selective Scanning (SS2D) module, which consists of three main components: the Cross-Scanning Block (Scan), the S6 Block, and the Cross-Merging Block (Merge).

As illustrated in Figure 3, the SS2D module consists of three main components: the Cross-Scanning Block (Scan), the S6 Block, and the Cross-Merging Block (Merge). The Cross-Scanning Block first scans the input feature map in four directions: left to right, top to bottom, right to left, and bottom to top, generating four sequential feature representations. Each sequence captures direction-specific contextual information, enabling a more comprehensive understanding of global dependencies within the feature map. These sequential features are then processed by the S6 Block for feature extraction. The S6 Block employs a state update mechanism, where the input sequence $x(t)$ is integrated with the hidden state $h(t)$ and mapped to the output $y(t)$ using the following equations:

$$h'(t) = \mathbf{A}h(t) + \mathbf{B}x(t),$$

$$y(t) = \mathbf{C}h(t),$$

where, $\mathbf{A} \in \mathbb{R}^{N\times N}$, $\mathbf{B} \in \mathbb{R}^{N\times 1}$, and $\mathbf{C} \in \mathbb{R}^{1\times N}$ are learnable parameter matrices. Here, $\mathbf{A}$ governs the temporal evolution of the hidden state $h(t)$, $\mathbf{B}$ determines the influence of the input $x(t)$ on the updated hidden state $h'(t)$, and $\mathbf{C}$ maps the hidden state $h(t)$ to the output $y(t)$.

Through this recurrent state-space formulation, the hidden state $h(t)$ integrates information from arbitrarily distant time steps, thereby enabling the model to capture long-range dependencies and encode global contextual information that extends beyond local neighborhoods. At the same time, the sequential update rules avoid explicit pairwise interactions between all tokens, unlike MHSA in Transformers. As a result, the computational cost of the S6 Block grows proportionally with sequence length, achieving near-linear complexity $O(N)$ rather than the quadratic complexity $O(N^2)$ of Transformers. This property makes the SS2D module particularly efficient and scalable for processing high-resolution medical images.

### 2.2.2 EFC-KAN Block

The EFC-KAN block builds on KAN, which parameterizes learnable basis functions using B-splines. The $k$-th order spline basis function at the $i$-th channel is computed as:

$$N_{i,k}(x_l) = \frac{x_l - t_i}{t_{i+k} - t_i} N_{i,k-1}(x_l) + \frac{t_{i+k+1} - x_l}{t_{i+k+1} - t_{i+1}} N_{i+1,k-1}(x_l),$$

where, $N_{i,k}(x_l)$ represents the $k$-th order basis function at the $i$-th channel, $t_i$ is the node at the $i$-th channel, and $x_l$ corresponds to the $l$-th pixel in the input feature map along the sequence dimension. However, because spline basis functions at each sequence position rely only on local input, spatial information exchange is limited, potentially hindering feature extraction. To address this, we introduce Expanded Field Convolution (EFConv) before the KAN block, forming the EFC-KAN block. EFConv comprises two 3×3 convolutional layers that expand the receptive field and enhance spatial information interaction.

The KAN block consists of two layers of KANLinear and DWConv, collectively referred to as the Tok-KAN block [30]. KANLinear, the core component of KAN, leverages B-spline functions to enable learnable weight representations. Its design is based on the Kolmogorov-Arnold theorem [28, 29], which states that any continuous function can be expressed as a composition of a finite number of continuous univariate functions. The KANLinear transformation is defined as:

$$\Phi(x) = \begin{pmatrix} \phi_{1,1}(\cdot) & \cdots & \phi_{1,n}(\cdot) \\ \vdots & \ddots & \vdots \\ \phi_{m,1}(\cdot) & \cdots & \phi_{m,n}(\cdot) \end{pmatrix} x,$$

where $\phi_{m,n}(\cdot)$ represents a learnable univariate function, and $m$ and $n$ denote the output and input feature dimensions. Each univariate function is parameterized as:

$$\phi_{m,n}(\cdot) = \phi(x) = \omega \cdot (b(x) + spline(x))$$

where $b(x)$ is a fixed base activation (SiLU), $spline(x)$ is a learnable B-spline function, and $\omega$ is a scaling factor. The spline coefficients are initialized using Kaiming uniform initialization to ensure stable optimization. To further mitigate overfitting, we adopt a pruning strategy on the B-spline coefficients, which reduces redundancy while preserving expressive capacity.

The full computation of the EFC-KAN block proceeds in three steps:

(1) Spatial expansion via EFConv:

$$\bar{x} = EFConv(x) = Conv(Conv(x)),$$

where two stacked 3×3 convolutions enhance spatial interactions.

(2) The EFConv output $\bar{x}$ is processed by the Tok-KAN block, formulated as:

$$Tok_{KAN(\bar{x})} = DWConv\left(KANLinear\left(DWConv\left(KANLinear\left(LN(\bar{x})\right)\right)\right)\right).$$

(3) The final output is obtained via a residual connection:

$$EFC_{KAN(\bar{x})} = Tok_{KAN(\bar{x})} + \bar{x}.$$

By combining EFConv with the Tok-KAN block, the EFC-KAN block achieves more effective spatial information exchange and enhanced nonlinear feature modeling. This design significantly improves the expressive power and generalization of the network, making it well-suited for complex medical image segmentation tasks.

### 2.3 CBAM Block

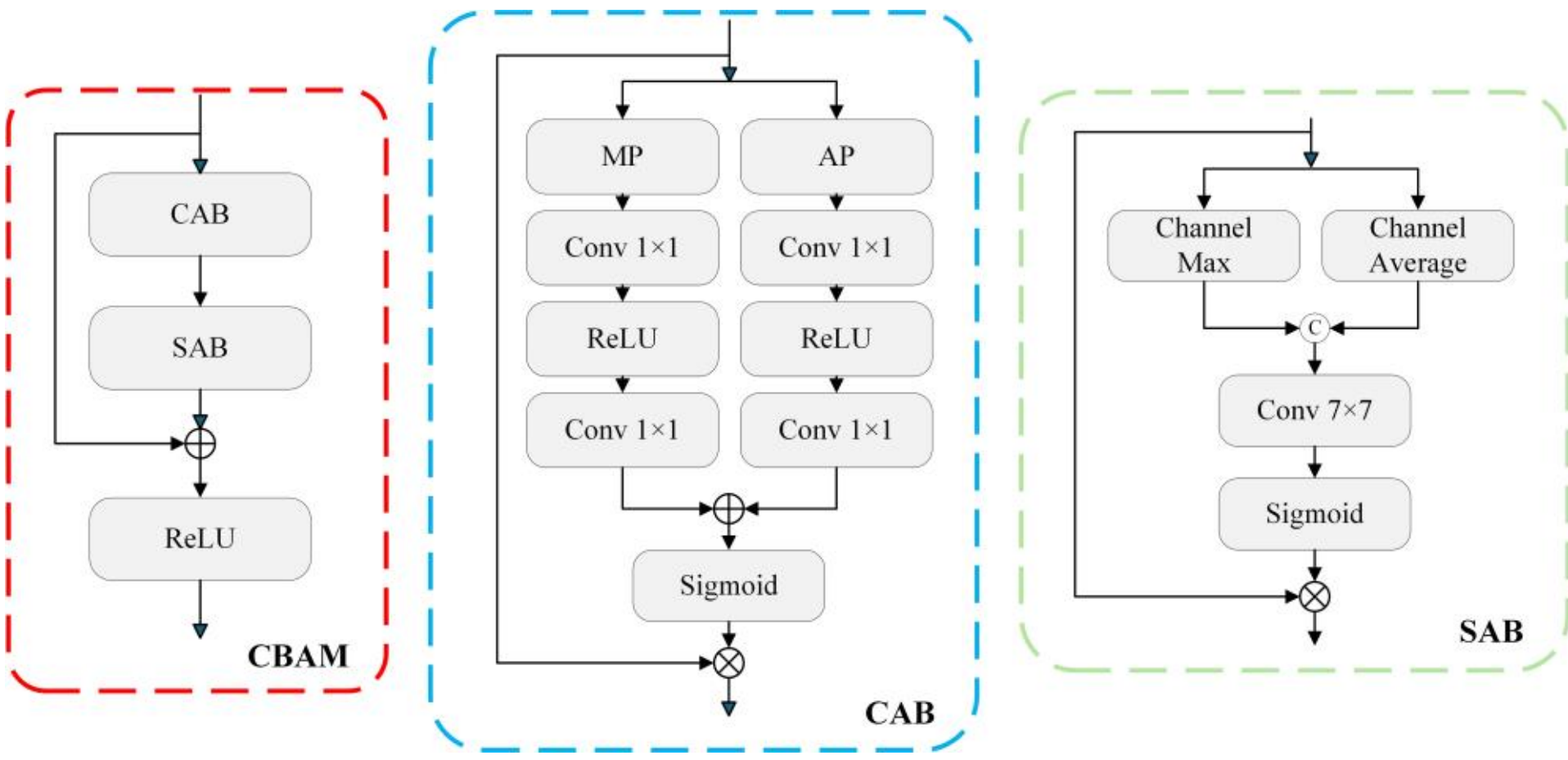

Figure 4. Schematic diagram of the CBAM structure, consisting of a Channel Attention Block (CAB) and a Spatial Attention Block (SAB).

The CBAM block enhances feature representation by sequentially refining the input feature map along the channel and spatial dimensions. As shown in Figure 4, the feature map first passes through the Channel Attention Block (CAB), which extracts channel-wise attention. CAB utilizes MP and average pooling (AP) along the spatial dimension in parallel branches to generate channel descriptors. These descriptors are processed by a $1 \times 1$ convolutiona for dimensionality reduction, followed by ReLU activation, and then another $1 \times 1$ convolutiona restores the original channel dimension. The outputs are summed and passed through a Sigmoid activation function to produce channel attention scores, which are applied via channel-wise multiplication to enhance the feature representation.

After channel refinement, the Spatial Attention Block (SAB) further enhances the feature map. It applies MP and AP along the channel dimension, concatenating the results to form a spatial descriptor. A 7×7 convolutional layer extracts spatial dependencies while reducing the channel dimension to 1. The output is then activated by Sigmoid, generating spatial attention scores, which are applied via element-wise multiplication to emphasize spatially significant features.

Finally, a residual connection adds the refined feature map to the original input, followed by ReLU activation, enhancing feature propagation and representational capacity.

### 2.4 Loss Function

Our network employs a deep supervision mechanism, where the total loss is a weighted sum of Dice loss and cross-entropy loss across multiple decoding stages:

$$Total\ Loss = \sum_{k=1}^{4} \alpha_k Loss_k \text{, with } Loss_k = DiceLoss_k + CrossEntropyLoss_k.$$

The Dice loss at stage $k$ is defined as:

$$DiceLoss_k = -\frac{2\left|Y \cap \tilde{Y}^k\right|}{|Y| + \left|\tilde{Y}^k\right|}.$$

The cross-entropy loss at stage $k$ is given by:

$$CrossEntropyLoss_k = -\frac{1}{N}\sum_{i=1}^{N}\sum_{j=1}^{C} y_{ij} \log p_{ij}^k.$$

In the above formulas, $Y$ represents the ground truth segmentation mask, and $\tilde{Y}^k$ denotes the predicted segmentation mask at stage $k$. $N$ is the total number of pixels, and $C$ is the number of segmentation classes. The term $y_{ij}$ is the one-hot encoded ground truth label for pixel $i$ in class $j$, while $p_{ij}^k$ represents the predicted probability of pixel $i$ belonging to class $j$ at stage $k$. The weighting coefficients $\alpha_1, \alpha_2, \alpha_3, \alpha_4$ are set to 1.0, 0.5, 0.25, and 0.125, respectively, corresponding to the four decoder stages. The first stage provides the final segmentation result, while the remaining stages contribute to feature refinement through deep supervision.

## 3 Experiments and Results

### 3.1 Datasets

We performed comprehensive experiments on five publicly available medical image segmentation datasets, covering a diverse range of segmentation tasks, including cell instance segmentation, multi-organ abdominal segmentation, breast cancer lesion segmentation, multi-class cardiac segmentation, and COVID-19 infection segmentation. The datasets encompass various imaging modalities, including microscopy, magnetic resonance imaging (MRI), ultrasound, and computed tomography (CT).

**The Microscopy dataset** originates from the NeurIPS 2022 Cell Segmentation Challenge [37], which focuses on cell segmentation across various microscopy images. Cell segmentation is a fundamental step in the quantitative analysis of individual cells in microscopic images. This dataset is specifically designed for instance segmentation, with images cropped to a resolution of (512, 512) for both training and testing. It comprises 1000 images for training and 101 images for testing. To ensure consistency, we adopted the same data preprocessing strategy as described in [21].

**The Abdomen MRI dataset** originates from the MICCAI 2022 AMOS Challenge [38], which focuses on segmenting 13 abdominal organs from MRI scans, including the liver, spleen, pancreas, right kidney, left kidney, stomach, gallbladder, esophagus, aorta, inferior vena cava, right adrenal gland, left adrenal gland, and duodenum. Following the experimental setup in [21], we incorporated an additional 50 MRI scans for testing. Specifically, the dataset comprises 60 MRI scans with 5615 slices for training and 50 MRI scans with 3357 slices for testing. All images were cropped to a resolution of (320, 320) for both training and testing.

**The BUSI dataset** was collected in 2018 by Walid Al-Dhabyani et al. from Cairo University, Egypt [39]. It comprises ultrasound images of normal, benign, and malignant breast cancer cases, along with their corresponding segmentation masks. For our study, we utilized all ultrasound images representing benign and malignant breast cancer cases, totaling 647 images. The images were resized to (512, 512) for both training and testing, with 517 images allocated for training and 130 for testing.

**The ACDC dataset** originates from the Automated Cardiac Diagnosis Challenge (ACDC) [40]. It consists of MRI scans from 150 patients, with each patient having two different MRI modalities. This challenge aims to segment the left ventricle, right ventricle, and myocardium from MRI scans. For our study, we utilized the training subset of the original dataset, which includes MRI scans from 100 patients. We further split this subset into training and testing sets using an 8:2 ratio, resulting in 160 MRI scans from 80 patients for training and 40 MRI scans from 20 patients for testing. All images were resized to (256, 256) for both training and testing.

**The COVID-19 dataset** was constructed by M$^2$SNet [41] through the integration of two publicly available collections: the COVID-19 CT Segmentation Dataset (https://medicalsegmentation.com/COVID19/) and the COVID-19 CT Scans Dataset from Zenodo (https://zenodo.org/records/3757476). Uniform sampling was applied to obtain a total of 1277 high-quality CT images, with 894 images designated for training and 383 for testing. All images were resized to (512, 512) for both training and testing.

### 3.2 Implementation Details

We trained MedVKAN using the nnU-Net framework [9], which automates hyperparameter configuration and data preprocessing. This automation enabled us to concentrate on model architecture design while retaining the flexibility to adjust hyperparameters when necessary. Moreover, using this framework ensured a fair comparison with other models under consistent experimental settings. All experiments were performed on a single NVIDIA GeForce RTX 4090D GPU. The initial learning rate was set to 0.0002 with a batch size of 8. We employed the AdamW optimizer with a weight decay of 0.05, and the learning rate followed a cosine annealing schedule with a minimum value of 0.000001. The model was trained for 1000 epochs.

For data preprocessing, we strictly followed the nnU-Net pipeline. Specifically, all images were cropped to nonzero regions to remove irrelevant background and retain clinically meaningful anatomical structures. The images were then resampled to the median voxel spacing of each dataset, with third-order spline interpolation applied to image data and nearest-neighbor interpolation applied to segmentation masks. For CT images, a two-stage normalization strategy was adopted: intensity values within the training masks were clipped to the [0.5, 99.5] percentile range, followed by z-score normalization using the mean and standard deviation of the dataset. For other imaging modalities, standard z-score normalization was directly applied. To further improve model generalization, various data augmentation techniques were incorporated during training, including random rotation, random scaling, elastic deformation, gamma correction, and mirroring.

### 3.3 Comparative Models and Evaluation Metrics

We conducted comparative experiments using eight representative models across six categories: traditional CNN-based methods (nnU-Net [9], SegResNet [13]); Transformer-based models (UNETR [15], SwinUNETR [16]); a CNN-Transformer hybrid model (nnW-Net[42]); an xLSTM-based model (xLSTM-UNet [43]); a Mamba-based model (Swin-UMamba [25]); and a KAN-based model (UKAN [30]). To ensure a fair comparison, all methods were implemented within the nnU-Net framework, with hyperparameters set according to their original publications. For consistency, all models were trained from scratch for 1,000 epochs under identical experimental settings.

Following [44], segmentation performance on the Abdomen MRI and ACDC datasets was evaluated using four metrics: Dice similarity coefficient, Jaccard index, normalized surface distance (NSD), and the 95th percentile Hausdorff distance (HD95). For the Microscopy dataset, which involves cell instance segmentation, three metrics were employed: F1-score, Jaccard index, and Dice similarity coefficient. For the BUSI and COVID-19 datasets, segmentation performance was assessed using the Jaccard index, Dice similarity coefficient, and HD95. The definitions of these metrics are provided below:

$$Dice = \frac{2|Y \cap \tilde{Y}|}{|Y| + |\tilde{Y}|},$$

$$NSD = \frac{|S_{pred} \cap S_{gt,\tau}| + |S_{gt} \cap S_{pred,\tau}|}{|S_{pred}| + |S_{gt}|},$$

$$Jaccard = \frac{|Y \cap \tilde{Y}|}{|Y \cup \tilde{Y}|},$$

$$F1 = 2 \times \frac{Precision \times Recall}{Precision + Recall}, \text{if Jaccard} > 0.5,$$

$$Precision = \frac{TP}{TP + FP},$$

$$Recall = \frac{TP}{TP + FN},$$

$$HD95(S_{pred}, S_{gt}) = max\left(h_{95}(S_{pred}, S_{gt}), h_{95}(S_{gt}, S_{pred})\right),$$

where

$$h_{95}(S_{pred}, S_{gt}) = percentile_{95}\left(\min_{y \in S_{gt}} d(x, y) | x \in S_{pred}\right).$$

In the above formulas, $Y$ represents the ground truth segmentation mask, while $\tilde{Y}$ denotes the predicted segmentation mask. $S_{pred}$ and $S_{gt}$ represent the surface point sets of the predicted and ground truth segmentations, respectively. Similarly, $S_{pred,\tau}$ and $S_{gt,\tau}$ refer to the sets of points lying within a distance threshold $\tau$ from the predicted and ground truth boundaries. The function $d(x, y)$ represents the Euclidean distance between point x and point y. For cell instance segmentation, a prediction is considered accurate if its Jaccard Index with the ground truth exceeds 0.5. Based on this criterion, the F1-score is computed, where true positives (TP) are correctly identified foreground pixels, false positives (FP) are background pixels incorrectly classified as foreground, and false negatives (FN) are foreground pixels incorrectly classified as background.

### 3.4 Experimental Results

Table 1. Experimental results of different models on the Microscopy dataset. The best and second-best results are highlighted in bold and underlined, respectively. Statistical significance is indicated by * (p < 0.01) and # (p < 0.05) based on the Wilcoxon signed-rank test.

| Methods | #Params | #FLOPs | Jaccard (mean±std) | Dice (mean±std) | F1 (mean±std) |
|---|---|---|---|---|---|
| nnU-Net | 46M | 59.0G | 40.2±24.4* | 69.9±18.2* | 52.9±26.2* |
| SegResNet | 6M | 62.4G | 41.3±25.3* | 68.6±22.7# | 53.8±27.1* |
| UNETR | 88M | 105.6G | 31.4±24.3* | **71.9±19.9** | 43.1±26.3* |
| SwinUNETR | 25M | 75.7G | 27.1±22.9* | 66.5±19.9* | 37.9±27.2* |
| nnW-Net | 7M | 42.2G | 43.3±27.4* | 68.0±23.0# | 55.0±28.8* |
| xLSTM-UNet | 86M | 101.2G | <u>44.8±23.5</u>* | <u>71.4±15.8</u> | <u>58.2±23.6</u> |
| UKAN | 25M | 27.4G | 41.2±26.8* | 66.7±22.2* | 52.9±28.9* |
| Swin-UMamba | 59M | 160.8G | 37.5±26.0* | 68.3±22.1* | 49.2±28.5* |
| MedVKAN | 51M | 56.5G | **48.4±26.7** | 70.0±20.5 | **60.4±27.4** |

Table 2. Experimental results of different models on the BUSI dataset. The best and second-best results are highlighted in bold and underlined, respectively. Statistical significance is indicated by * (p < 0.01) and # (p < 0.05) based on the Wilcoxon signed-rank test.

| Methods | #Params | #FLOPs | Jaccard (mean±std) | Dice (mean±std) | HD95 (mean±std) |
|---|---|---|---|---|---|
| nnU-Net | 46M | 59.0G | 68.0±28.4* | 76.3±27.6* | 59.9±117.5* |
| SegResNet | 6M | 62.4G | 65.3±28.6* | 74.3±27.3* | 87.8±137.4* |
| UNETR | 88M | 105.6G | 59.7±29.6* | 69.5±28.9* | 125.7±150.5* |
| SwinUNETR | 25M | 75.7G | 67.1±27.4* | 76.2±25.9* | 76.6±123.0* |
| nnW-Net | 7M | 42.2G | 68.3±27.3* | 77.0±25.8* | 57.9±112.8* |
| xLSTM-UNet | 86M | 101.2G | 69.2±28.6* | 77.1±28.0* | <u>51.1±109.3</u>* |
| UKAN | 25M | 27.4G | 66.4±28.8* | 75.0±28.1* | 65.1±120.9* |
| Swin-UMamba | 59M | 160.8G | <u>69.9±26.8</u># | <u>78.3±25.4</u># | 52.1±100.5* |
| MedVKAN | 51M | 56.5G | **70.9±27.3** | **78.8±26.4** | **48.9±108.7** |

As shown in Table 1, although MedVKAN reports slightly lower pixel-level Dice scores than UNETR and xLSTM-UNet, it achieves the highest instance-level Jaccard Index and F1-score, surpassing the second-best model by 3.6% and 2.2%, respectively. Since the Microscopy dataset involves cell instance segmentation, where segmentation completeness is crucial, the F1-score is computed only for instances with Jaccard > 0.5. A higher F1-score therefore indicates that MedVKAN successfully segments a larger number of cell instances with sufficient accuracy. This demonstrates that, despite slightly lower pixel-level overlap, MedVKAN achieves superior instance-level segmentation completeness and recall, both of which are critical for reliable cell analysis.

On the BUSI dataset (table 2), MedVKAN consistently outperforms Swin-UMamba, the second-best model, achieving 1.0% higher Jaccard and 0.5% higher Dice scores, while also maintaining the lowest HD95. Similarly, on the Abdomen MRI dataset (table 3), MedVKAN surpasses the second-best model by 1.3% in Jaccard, 0.8% in Dice, and 1.0% in NSD, while again achieving the lowest HD95. These improvements indicate that MedVKAN delivers more accurate segmentation and exhibits stronger robustness in boundary delineation.

Table 3. Experimental results of different models on the Abdomen MRI dataset. The best and second-best results are highlighted in bold and underlined, respectively. Statistical significance is indicated by * ($p < 0.01$) and # ($p < 0.05$) based on the Wilcoxon signed-rank test.

| Methods | #Params | #FLOPs | Jaccard (mean±std) | Dice (mean±std) | HD95 (mean±std) | NSD (mean±std) |
|---|---|---|---|---|---|---|
| nnU-Net | 33M | 23.0G | 64.7±11.3* | 74.7±10.9* | 21.2±15.6* | 81.5±11.4* |
| SegResNet | 6M | 24.3G | 65.0±13.6* | 74.7±13.2* | 23.5±16.8* | 81.5±13.0* |
| UNETR | 87M | 41.0G | 48.2±15.3* | 58.7±16.4* | 45.4±34.8* | 64.1±17.9* |
| SwinUNETR | 25M | 29.4G | 60.0±13.5* | 70.1±13.7* | 27.1±21.7* | 76.4±14.9* |
| nnW-Net | 7M | 16.5G | <u>65.5±12.3</u>* | 75.2±11.7* | 21.5±17.6* | 81.2±12.2* |
| xLSTM-UNet | 63M | 41.1G | <u>65.5±9.7</u>* | 73.8±8.9* | 26.7±11.8* | 78.9±8.9* |
| UKAN | 25M | 10.7G | 63.1±14.2* | 73.0±13.9* | 23.9±17.5* | 79.7±14.0* |
| Swin-UMamba | 59M | 62.7G | <u>65.5±11.7</u>* | <u>75.4±11.3</u>* | <u>19.8±14.9</u># | <u>82.0±11.5</u>* |
| MedVKAN | 51M | 22.0G | **66.8±12.1** | **76.2±11.4** | **18.2±14.0** | **83.0±11.2** |

Table 4. Experimental results of different models on the ACDC dataset. The best and second-best results are highlighted in bold and underlined, respectively. Statistical significance is indicated by * ($p < 0.01$) and # ($p < 0.05$) based on the Wilcoxon signed-rank test.

| Methods | #Params | #FLOPs | Jaccard (mean±std) | Dice (mean±std) | HD95 (mean±std) | NSD (mean±std) |
|---|---|---|---|---|---|---|
| nnU-Net | 20M | 14.7G | 84.4±5.0* | 91.3±3.1* | 4.5±4.3# | 97.4±3.0# |
| SegResNet | 6M | 15.5G | 85.2±4.6* | 91.6±3.0* | 4.8±7.5 | <u>98.0±2.6</u> |
| UNETR | 87M | 26.2G | 78.0±7.6* | 86.9±5.6* | 11.4±18.9* | 93.8±6.3* |
| SwinUNETR | 25M | 19.0G | 83.6±4.9* | 90.8±3.1* | 5.4±5.9* | 97.1±2.8* |
| nnW-Net | 7M | 8.1G | <u>85.5±5.3</u> | <u>91.8±3.6</u> | <u>3.5±2.5</u> | 97.9±2.9 |
| xLSTM-UNet | 39M | 24.2G | 84.9±4.9* | 91.6±3.2* | **3.4±2.5** | 97.8±2.8 |
| UKAN | 25M | 6.8G | 84.5±5.3* | 91.3±3.7* | 3.6±2.6 | 97.7±3.0 |
| Swin-UMamba | 59M | 40.0G | 84.4±5.6* | 91.2±3.8* | 6.5±10.6* | 97.1±3.7* |
| MedVKAN | 51M | 14.0G | **85.9±4.2** | **92.2±2.7** | 4.0±4.3 | **98.1±2.1** |

Table 4 highlights MedVKAN's strong performance on the ACDC dataset, where all models achieve relatively high accuracy. Although MedVKAN does not yield the lowest HD95, it surpasses the second-best model by 0.4% in Jaccard, 0.4% in Dice, and 0.1% in NSD, demonstrating robustness in cardiac segmentation. On the COVID-19 dataset (Table 5), MedVKAN ranks second in Jaccard and Dice but maintains a clear advantage over most competing models and achieves the lowest HD95, reaffirming its effectiveness in infection region segmentation. Notably, xLSTM-UNet failed to converge due to training instability, and its results are therefore not reported.

Statistical analyses with the Wilcoxon signed-rank test further confirm the effectiveness of MedVKAN. Across five benchmark datasets, MedVKAN achieves significant improvements over most baseline methods on key metrics including Jaccard, Dice, F1-score, NSD and HD95, underscoring its consistent advantage in accurate and reliable segmentation. While MedVKAN does not have the smallest parameter count, it delivers substantially lower floating-point operations (FLOPs) than Transformer-based models, demonstrating a favorable balance between efficiency and accuracy.

Table 5. Experimental results of different models on the COVID-19 dataset. The best and second-best results are highlighted in bold and underlined, respectively. Statistical significance is indicated by * ($p < 0.01$) and # ($p < 0.05$) based on the Wilcoxon signed-rank test.

| Methods | #Params | #FLOPs | Jaccard (mean±std) | Dice (mean±std) | HD95 (mean±std) |
|---|---|---|---|---|---|
| nnU-Net | 46M | 59.0G | 62.0±35.2# | 68.8±36.0# | 29.1±62.8* |
| SegResNet | 6M | 62.2G | 61.2±34.9* | 68.2±35.9* | 41.0±75.6* |
| UNETR | 87M | 105.1G | 59.6±35.6* | 66.7±36.6* | 41.3±80.1* |
| SwinUNETR | 25M | 75.4G | 61.2±35.5* | 68.1±36.3* | 40.7±83.3* |
| nnW-Net | 7M | 42.1G | 61.9±35.7# | 68.6±36.3# | <u>27.5±64.6</u> |
| UKAN | 25M | 27.2G | 60.2±36.4# | 66.8±37.2# | 30.3±68.4# |
| Swin-UMamba | 59M | 160.2G | **63.5±34.9** | **70.1±35.7** | 31.8±76.7* |
| MedVKAN | 51M | 56.2G | <u>63.0±35.2</u> | <u>69.6±36.0</u> | **27.4±64.4** |

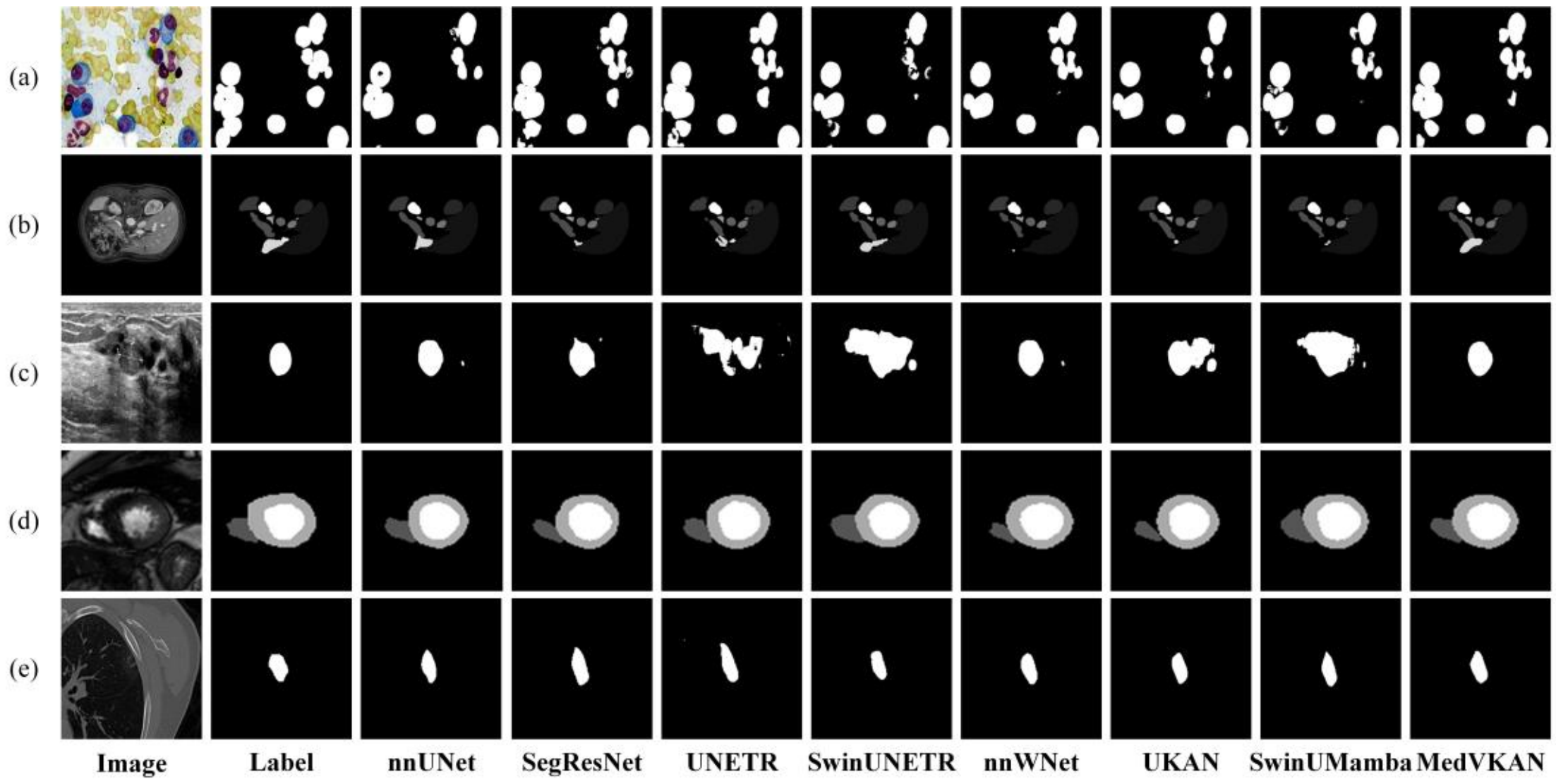

Figure 5. Visualization of segmentation results from different models. Rows (a)-(e) correspond to the Microscopy, Abdomen MRI, BUSI, ACDC, and COVID-19 datasets, respectively. For better visibility, images in rows (d) and (e) were resized and cropped due to the small size of the segmentation regions.

Figure 5 presents a visual comparison of segmentation results across five datasets. On the Microscopy dataset (row a), Swin-UMamba and UKAN fail to clearly separate adjacent or overlapping cells, leading to fragmented boundaries, whereas MedVKAN achieves more complete and coherent cell segmentations. On the Abdomen MRI dataset (row b), MedVKAN demonstrates superior capability in delineating closely overlapping organ boundaries, producing results that are both more accurate and structurally consistent with the ground truth. On the BUSI dataset (row c), MedVKAN generates smoother and cleaner breast lesion contours, while other models introduce irregular noise along the boundaries.

On the ACDC dataset (row d) and the COVID-19 dataset (row e), all models produce reasonably good results. However, MedVKAN achieves segmentations that are visually closer to the ground truth, providing more anatomically faithful delineation of the left ventricle, right ventricle, and myocardium in ACDC, as well as more complete and smoother delineation of infection regions in COVID-19. These observations highlight MedVKAN's ability to deliver highly reliable and realistic segmentation outcomes across diverse tasks and imaging modalities.

### 3.5 Ablation Study

Our VKAN block adopts an architecture analogous to the Transformer block. To assess its effectiveness, we performed an ablation study comparing it with the Transformer baseline, as summarized in Table 6. The variants include using only the EFC-KAN block, only the VSS block, the standard Transformer configuration (MHSA+MLP), and hybrid designs where either MHSA or MLP is replaced by VSS or EFC-KAN. Among these configurations, the proposed VSS+EFC-KAN model achieves the highest Jaccard, Dice, and HD95 performance, demonstrating that Mamba and KAN provide complementary benefits and jointly enhance segmentation accuracy.

To evaluate the influence of convolutional kernel size in EFConv on EFC-KAN's feature extraction, we conducted an additional ablation study (Table 7). In this study, "Conv3×3" denotes a single 3×3 convolution, "Conv5×5" a single 5×5 convolution, and "2×Conv3×3" two stacked 3×3 convolutions that provide an effective receptive field comparable to "Conv5×5". The results demonstrate that incorporating Conv3×3 already yields substantial improvements in Jaccard and Dice over the baseline. Using Conv5×5 provides marginal additional gains. Notably, 2×Conv3×3 achieves the highest Jaccard and Dice scores, slightly surpassing Conv5×5, while requiring fewer parameters and lower computational cost, making it the most efficient configuration.

Table 6. Ablation study of VKAN block variants on the BUSI dataset. The best results are highlighted in bold.

| Components | | | | Metrics | | |
|---|---|---|---|---|---|---|
| VSS | MHSA | EFC-KAN | MLP | Jaccard | Dice | HD95 |
| × | × | √ | × | 65.6±27.7 | 74.5±28.8 | 67.7±115.2 |
| √ | × | × | × | 68.4±28.2 | 76.7±27.5 | 52.6±99.1 |
| × | √ | × | √ | 60.7±29.8 | 70.2±29.2 | 105.7±141.9 |
| √ | × | × | √ | 69.0±29.1 | 76.8±28.5 | 51.1±110.0 |
| × | √ | √ | × | 66.3±27.0 | 75.9±24.4 | 90.5±123.7 |
| √ | × | √ | × | **70.9±27.3** | **78.8±26.4** | **48.9±108.7** |

Table 7. Ablation study of kernel configurations in EFConv on the BUSI dataset. Best results are highlighted in bold.

| Components | | | Metrics | | |
|---|---|---|---|---|---|
| Conv3 × 3 | Conv5 × 5 | 2 ∗ Conv3 × 3 | Jaccard | Dice | HD95 |
| × | × | × | 67.7±29.5 | 75.7±29.0 | 51.4±100.1 |
| √ | × | × | 69.8±28.2 | 77.7±27.6 | **38.8±72.9** |
| × | √ | × | 70.2±27.3 | 78.3±26.3 | 48.8±94.6 |
| × | × | √ | **70.9±27.3** | **78.8±26.4** | 48.9±108.7 |

## 4 Discussion

### 4.1 Computational Efficiency Analysis

The computational efficiency of MedVKAN was evaluated by comparing its training cost with that of other representative models, with particular focus on Transformer-based architectures such as UNETR and SwinUNETR. The evaluation was based on training memory usage, training time per epoch, and inference time on the Abdomen MRI dataset. As reported in Table 8, MedVKAN achieves memory consumption (7.61 GB) and training time (48 s/epoch) comparable to CNN-based models (e.g., nnU-Net: 7.77 GB, 52 s/epoch), while being substantially more efficient than Transformer-based models (e.g., UNETR: 18.39 GB, 102 s/epoch; SwinUNETR: 18.39 GB, 94 s/epoch). Although its inference time (3.18 s) is slightly higher than that of some CNN-based models, MedVKAN remains competitive and considerably faster than UNETR, SwinUNETR, and Swin-UMamba. Overall, these findings demonstrate that MedVKAN achieves a favorable trade-off between segmentation accuracy and computational cost, offering CNN-level efficiency while retaining the representational capacity of advanced sequence models.

Table 8. Comparison of training memory, training time per epoch, and inference time for different models on the Abdomen MRI dataset.

| Methods | Training Memory (GB) | Training Time (s) | Inference Time (s) |
|---|---|---|---|
| nnU-Net | 7.77 | 52 | 2.91 |
| SegResNet | 11.94 | 59 | 2.91 |
| UNETR | 18.39 | 102 | 3.22 |
| SwinUNETR | 18.39 | 94 | 3.30 |
| nnW-Net | 15.14 | 75 | 3.04 |
| xLSTM-UNet | 10.63 | 112 | 2.93 |
| UKAN | 6.80 | 37 | 3.01 |
| Swin-UMamba | 18.02 | 169 | 3.52 |
| MedVKAN | 7.61 | 48 | 3.18 |

Table 9. Dice scores on the test set for different activation functions (GELU, ReLU, SiLU, and KAN) on the ACDC dataset.

| Method | GELU | ReLU | SiLU | KAN |
|---|---|---|---|---|
| Dice | 91.9 | 91.7 | 91.9 | 92.2 |

### 4.2 Effectiveness of Learnable Activation Functions in KAN

To investigate the effectiveness of learnable activation functions in KAN, we conducted controlled experiments on the ACDC dataset by replacing KAN's adaptive activation with three widely used fixed functions: GELU, ReLU, and SiLU. Figure 6 shows that models with fixed activations converge more slowly, with losses decreasing gradually across epochs, whereas the KAN model converges substantially faster, suggesting that adaptive activations promote more efficient gradient flow and optimization.

Beyond training efficiency, Table 9 reports the Dice scores on the test set under different activation functions. While all fixed activations achieve competitive results, KAN consistently outperforms them, reaching the highest Dice score of 92.2%. This improvement highlights superior generalization capacity, as the learnable activation can flexibly adapt to complex data variations that fixed functions may fail to capture.

Overall, these results confirm that learnable activations in KAN are not just a theoretical refinement but provide tangible benefits: they speed up convergence, stabilize training, and improve generalization in challenging scenarios, establishing them as a key architectural component for advancing medical image segmentation models.

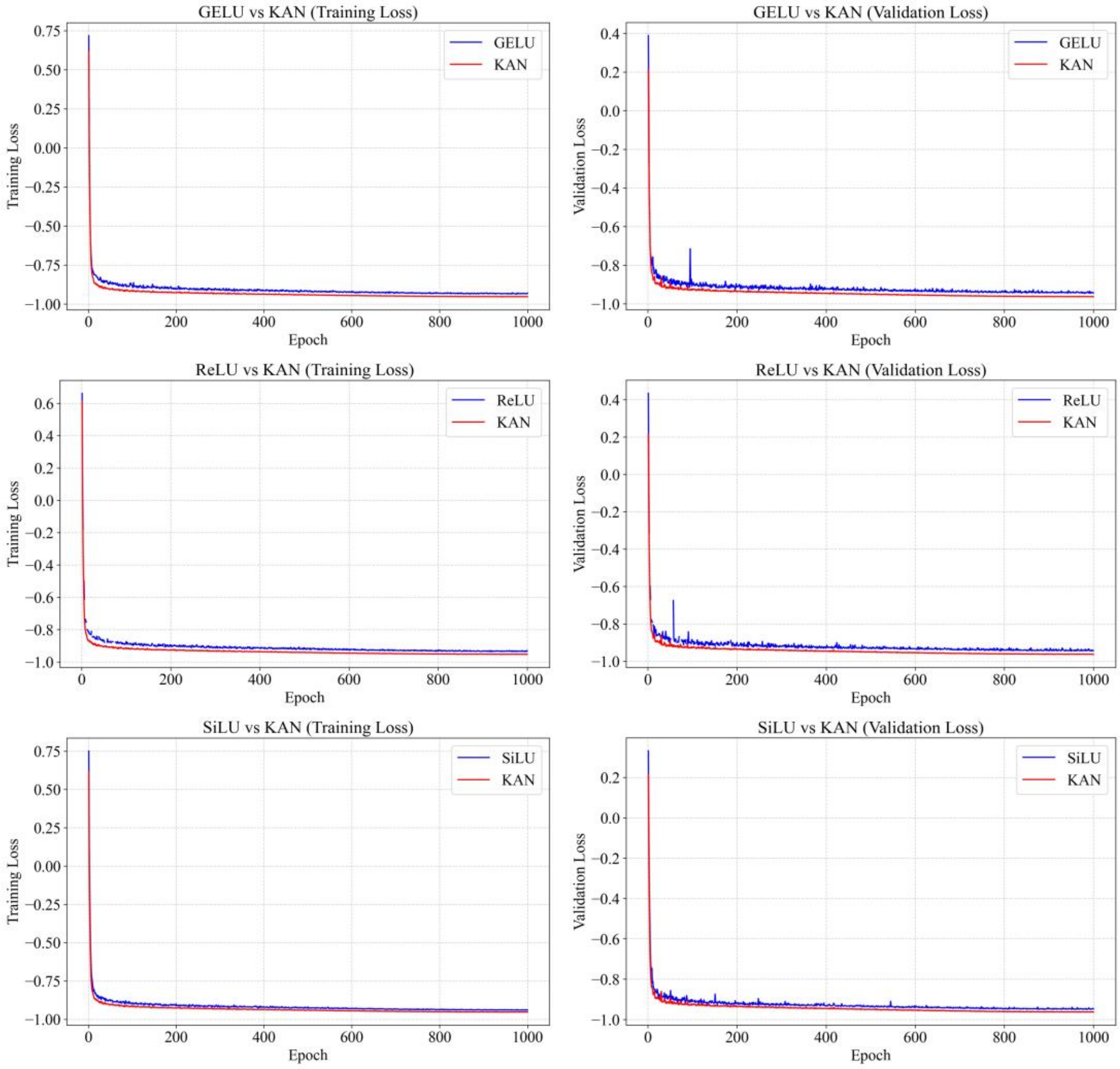

Figure 6. Training and validation loss curves for different activation functions (KAN, GELU, ReLU, and SiLU) on the ACDC dataset.

### 4.3 Limitations and Future Work

While MedVKAN demonstrates the benefits of integrating Mamba and KAN, its improvements over strong baselines are not always substantial. One reason is that the VSS block, although effective for modeling long-range dependencies, may not fully capture cross-scale spatial interactions in datasets with complex structures. Similarly, the EFC-KAN block enhances nonlinear expressiveness, but its effect may be limited by redundancy in feature representations, thereby constraining further accuracy gains. In addition, when datasets are relatively saturated and most models already achieve high accuracy, performance improvements naturally appear marginal, reflecting a ceiling effect rather than a lack of methodological contribution.

Future work will focus on enhancing the representational capacity of MedVKAN. Promising directions include developing multi-scale and multimodal fusion strategies to capture complementary information, designing adaptive mechanisms to balance local detail modeling with global context learning, and incorporating domain-specific priors to improve generalization and robustness across diverse clinical applications. With these refinements, MedVKAN has the potential to achieve greater accuracy gains while preserving efficiency, offering a reliable and scalable solution for medical image segmentation.

## 5 Conclusion

This study presents MedVKAN, a medical image segmentation model that integrates U-Net's strong local feature extraction with Mamba's efficient global sequence modeling, further enhanced by the EFC-KAN module for nonlinear feature representation. The proposed framework represents an initial exploration of the joint application of Mamba and KAN in medical image analysis. Extensive experiments across multiple datasets demonstrate that MedVKAN achieves competitive performance on diverse modalities and tasks. Moreover, the ablation studies confirm the complementary strengths of Mamba and KAN, highlighting the effectiveness of their integration in improving both optimization efficiency and segmentation accuracy.

### CRediT authorship contribution statement

Hancan Zhu: Writing-original draft, Investigation, Supervision. Jinhao Chen: Methodology, Conceptualization, Software. Guanghua He: Writing-review & editing, Methodology, Supervision.

### Declaration of competing interest

The authors declare that they have no competing financial interests.

### Acknowledgment

This work was supported by the Humanities and Social Science Fund of Ministry of Education of China (23YJAZH232) and the Scientific Research Project of Shaoxing University (20210038).

### Data availability

All datasets utilized in this study are publicly accessible.

## References

[1] X. Mei *et al.*, "Artificial intelligence–enabled rapid diagnosis of patients with COVID-19," *Nature medicine,* vol. 26, no. 8, pp. 1224-1228, 2020.

[2] L. Dai *et al.*, "A deep learning system for predicting time to progression of diabetic retinopathy," *Nature Medicine,* vol. 30, no. 2, pp. 584-594, 2024.

[3] X. Chen *et al.*, "Recent advances and clinical applications of deep learning in medical image analysis," *Medical image analysis,* vol. 79, p. 102444, 2022.

[4] Y. LeCun, L. Bottou, Y. Bengio, and P. Haffner, "Gradient-based learning applied to document recognition," *Proceedings of the IEEE,* vol. 86, no. 11, pp. 2278-2324, 1998.

[5] A. Jungo *et al.*, "On the effect of inter-observer variability for a reliable estimation of uncertainty of medical image segmentation," in *Medical Image Computing and Computer Assisted Intervention–MICCAI 2018: 21st International Conference, Granada, Spain, September 16-20, 2018, Proceedings, Part I*, 2018: Springer, pp. 682-690.

[6] M. Khened, V. A. Kollerathu, and G. Krishnamurthi, "Fully convolutional multi-scale residual DenseNets for cardiac segmentation and automated cardiac diagnosis using ensemble of classifiers," *Medical image analysis,* vol. 51, pp. 21-45, 2019.

[7] X. Jiang *et al.*, "End-to-end prognostication in colorectal cancer by deep learning: a retrospective, multicentre study," *The Lancet Digital Health,* vol. 6, no. 1, pp. e33-e43, 2024.

[8] O. Ronneberger, P. Fischer, and T. Brox, "U-net: Convolutional networks for biomedical image segmentation," in *Medical image computing and computer-assisted intervention–MICCAI 2015:*

*18th international conference, Munich, Germany, October 5-9, 2015, proceedings, part III 18*, 2015: Springer, pp. 234-241.

[9] F. Isensee, P. F. Jaeger, S. A. Kohl, J. Petersen, and K. H. Maier-Hein, "nnU-Net: a self-configuring method for deep learning-based biomedical image segmentation," *Nature methods,* vol. 18, no. 2, pp. 203-211, 2021.

[10] A. Vaswani *et al.*, "Attention is all you need," *Advances in neural information processing systems,* vol. 30, pp. 5998–6008, 2017.

[11] A. Dosovitskiy *et al.*, "An image is worth 16x16 words: Transformers for image recognition at scale," *arXiv preprint arXiv:2010.11929,* 2020.

[12] K. He, X. Zhang, S. Ren, and J. Sun, "Deep residual learning for image recognition," in *Proceedings of the IEEE conference on computer vision and pattern recognition*, 2016, pp. 770-778.

[13] A. Myronenko, "3D MRI brain tumor segmentation using autoencoder regularization," in *Brainlesion: Glioma, Multiple Sclerosis, Stroke and Traumatic Brain Injuries: 4th International Workshop, BrainLes 2018, Held in Conjunction with MICCAI 2018, Granada, Spain, September 16, 2018, Revised Selected Papers, Part II 4*, 2019: Springer, pp. 311-320.

[14] Z. Liu *et al.*, "Swin transformer: Hierarchical vision transformer using shifted windows," in *Proceedings of the IEEE/CVF international conference on computer vision*, 2021, pp. 10012-10022.

[15] A. Hatamizadeh *et al.*, "Unetr: Transformers for 3d medical image segmentation," in *Proceedings of the IEEE/CVF winter conference on applications of computer vision*, 2022, pp. 574-584.

[16] A. Hatamizadeh, V. Nath, Y. Tang, D. Yang, H. R. Roth, and D. Xu, "Swin unetr: Swin transformers for semantic segmentation of brain tumors in mri images," in *International MICCAI brainlesion workshop*, 2021: Springer, pp. 272-284.

[17] A. Gu, K. Goel, and C. Ré, "Efficiently modeling long sequences with structured state spaces," *arXiv preprint arXiv:2111.00396,* 2021.

[18] A. Gu *et al.*, "Combining recurrent, convolutional, and continuous-time models with linear state space layers," *Advances in neural information processing systems,* vol. 34, pp. 572-585, 2021.

[19] A. Gu and T. Dao, "Mamba: Linear-time sequence modeling with selective state spaces," *arXiv preprint arXiv:2312.00752,* 2023.

[20] B. L. Lianghui Zhu, Qian Zhang, Xinlong Wang, Wenyu Liu, Xinggang Wang, "Vision Mamba: Efficient Visual Representation Learning with Bidirectional State Space Model," *Forty-first International Conference on Machine Learning,* 2024.

[21] J. Ma, F. Li, and B. Wang, "U-mamba: Enhancing long-range dependency for biomedical image segmentation," *arXiv preprint arXiv:2401.04722,* 2024.

[22] Y. Liu *et al.*, "Vmamba: Visual state space model," *Advances in neural information processing systems,* vol. 37, pp. 103031-103063, 2024.

[23] Z. Han, C. Gao, J. Liu, J. Zhang, and S. Q. Zhang, "Parameter-efficient fine-tuning for large models: A comprehensive survey," *arXiv preprint arXiv:2403.14608,* 2024.

[24] Z. Xing, T. Ye, Y. Yang, G. Liu, and L. Zhu, "Segmamba: Long-range sequential modeling mamba for 3d medical image segmentation," in *International Conference on Medical Image Computing and Computer-Assisted Intervention*, 2024: Springer, pp. 578-588.

[25] J. Liu *et al.*, "Swin-umamba: Mamba-based unet with imagenet-based pretraining," in *International Conference on Medical Image Computing and Computer-Assisted Intervention*, 2024: Springer, pp. 615-625.

[26] Z. Liu *et al.*, "Kan: Kolmogorov-arnold networks," *arXiv preprint arXiv:2404.19756,* 2024.

[27] Z. Liu, P. Ma, Y. Wang, W. Matusik, and M. Tegmark, "Kan 2.0: Kolmogorov-arnold networks meet science," *arXiv preprint arXiv:2408.10205,* 2024.

[28] A. N. Kolmogorov, *On the representation of continuous functions of several variables by superpositions of continuous functions of a smaller number of variables*. American Mathematical Society, 1961.

[29] V. Tikhomirov, "On the representation of continuous functions of several variables as superpositions of continuous functions of one variable and addition," in *Selected Works of AN Kolmogorov*: Springer, 1991, pp. 383-387.

[30] C. Li *et al.*, "U-kan makes strong backbone for medical image segmentation and generation," *arXiv preprint arXiv:2406.02918,* 2024.

[31] X. Yang and X. Wang, "Kolmogorov-arnold transformer," *arXiv preprint arXiv:2409.10594,* 2024.

[32] A. D. Bodner, A. S. Tepsich, J. N. Spolski, and S. Pourteau, "Convolutional kolmogorov-arnold networks," *arXiv preprint arXiv:2406.13155,* 2024.

[33] Y. Wu, T. Li, Z. Wang, H. Kang, and A. He, "TransUKAN: Computing-Efficient Hybrid KAN-Transformer for Enhanced Medical Image Segmentation," *arXiv preprint arXiv:2409.14676,* 2024.

[34] C.-Y. Lee, S. Xie, P. Gallagher, Z. Zhang, and Z. Tu, "Deeply-supervised nets," in *Artificial intelligence and statistics*, 2015: Pmlr, pp. 562-570.

[35] S. Woo, J. Park, J.-Y. Lee, and I. S. Kweon, "Cbam: Convolutional block attention module," in *Proceedings of the European conference on computer vision (ECCV)*, 2018, pp. 3-19.

[36] H. Cao *et al.*, "Swin-unet: Unet-like pure transformer for medical image segmentation," in *European conference on computer vision*, 2022: Springer, pp. 205-218.

[37] J. Ma *et al.*, "The multimodality cell segmentation challenge: toward universal solutions," *Nature methods,* pp. 1-11, 2024.

[38] J. Ma *et al.*, "Unleashing the strengths of unlabeled data in pan-cancer abdominal organ quantification: the flare22 challenge," *arXiv preprint arXiv:2308.05862,* 2023.

[39] W. Al-Dhabyani, M. Gomaa, H. Khaled, and A. Fahmy, "Dataset of breast ultrasound images," *Data in brief,* vol. 28, p. 104863, 2020.

[40] O. Bernard *et al.*, "Deep learning techniques for automatic MRI cardiac multi-structures segmentation and diagnosis: is the problem solved?," *IEEE transactions on medical imaging,* vol. 37, no. 11, pp. 2514-2525, 2018.

[41] X. Zhao *et al.*, "M2SNet: Multi-scale in multi-scale subtraction network for medical image segmentation," *arXiv preprint arXiv:2303.10894,* 2023.

[42] Y. Zhou, L. Li, L. Lu, and M. Xu, "nnWNet: Rethinking the Use of Transformers in Biomedical Image Segmentation and Calling for a Unified Evaluation Benchmark," in *Proceedings of the Computer Vision and Pattern Recognition Conference*, 2025, pp. 20852-20862.

[43] T. Chen *et al.*, "xLSTM-UNet can be an Effective Backbone for 2D & 3D Biomedical Image Segmentation Better than its Mamba Counterparts," in *2024 IEEE EMBS International Conference on Biomedical and Health Informatics (BHI)*, 2024: IEEE, pp. 1-8.

[44] L. Maier-Hein and B. Menze, "Metrics reloaded: Pitfalls and recommendations for image analysis validation," *arXiv. org,* no. 2206.01653, 2022.